\documentclass[conference]{IEEEtran}
\usepackage{graphicx}
\usepackage{url}
\usepackage[linesnumbered,ruled,vlined]{algorithm2e}

\usepackage{cite}

\ifCLASSINFOpdf
\else
\fi
\begin{document}
%

\title{A Novel Software Tool for Analysing  NT\textsuperscript{\textregistered}  File System Permissions}

\author{\IEEEauthorblockN{Simon Parkinson and Andrew Crampton}
\IEEEauthorblockA{School of Informatics\\
University of Huddersfield\\
HD1 3DH, UK\\
Email: s.parkinson@hud.ac.uk}}


%


\maketitle

\begin{abstract}
Administrating and monitoring New Technology File System (NTFS) permissions can be a cumbersome and convoluted task. In today's data rich world there has never been a more important time to ensure that data is secured against unwanted access. This paper identifies the essential and fundamental requirements of access control, highlighting the main causes of their misconfiguration within the NTFS. In response, a number of features are identified and an efficient, informative and intuitive software-based solution is proposed for examining file system permissions. In the first year that the software has been made freely available it has been downloaded and installed by over four thousand users\footnote{Available at: 
http://eprints.hud.ac.uk\/9743\/  \\
and
http://download.cnet.com\/NTFS\-Permissions\-Explorer\-SnapIn\/3000\-2094\_4\-75325639}.
\end{abstract}


%
\IEEEpeerreviewmaketitle

\section{Introduction}
Controlling access permissions to a given file system is an important aspect of data security. Having a secure and flexible way of viewing and managing access control should be a standard requirement of all modern file systems. This should certainly be true of the New Technology File System (NTFS), since NTFS is currently the most common file system in use. This is mainly due to Microsoft's dominance of computing operating systems. Surprisingly, however, no such flexibility exists for the NTFS and the process for determining access controls is cumbersome at best.

The NTFS  implements access control with the use of Access Control Lists (ACLs). Each file system object (folder or file) will have an associated ACL for controlling access. An ACL contains a list of ACEs (Access Control Entities). Each ACE contains information regarding the interacting user or group, and the level of access that they will be granted. 

It is well reported that from observing an ACE that the following information can be established \cite{russel2003, rusinovich2005, admincomp}:
\begin{enumerate}
	\item The user or group that the ACE applies to.
	\item The level of granted permission for a user or group.
	\item Information regarding the prorogation of the permission down the directory hierarchy 
\end{enumerate}

\begin{figure}
\centering
\includegraphics[width=2.5in]{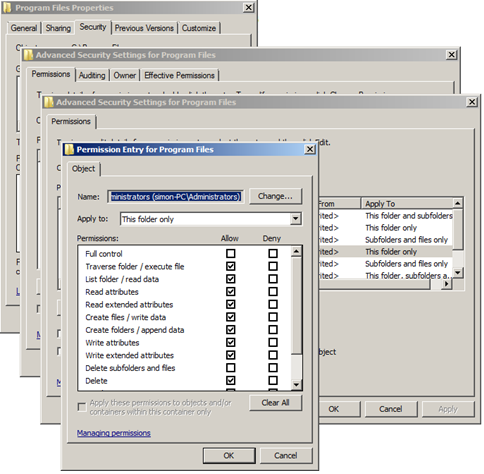}
\caption{Analysing NTFS file system permissions using Windows Explorer}
\label{fig:explorerclicks}
\end{figure}

The way in which users are required to interact with ACEs and ACLs in the NTFS results in the following peculiarities:

\begin{enumerate}
	\item Permissions are interacted with on a per object level, rather than per user \cite{Hanner:1999fk}. This does not allow for the administrator to evaluate user permission across a whole directory structure.
	\item Interacting with a single ACL using Windows Explorer as seen in Figure~\ref{fig:explorerclicks} requires the traversal of four different interfaces. Interacting with multiple ACLs soon becomes a cumbersome task, which could ultimately result in permissions being overlooked.
	\item Not only is the administrator required to examine users or groups within the ACL, they have to remember, or explore, group association to evaluate the inheritance of permissions from different groups. 
\end{enumerate}
It is well reported that these time-consuming peculiarities result in the potential for errors to occur, which could ultimately result in users being denied access, or in the worst case, the possibility for unwanted access to occur \cite{Beznosov:2009, Cao:2006, Maxion2005, Hanner:1999fk, admincomp}.

Previous efforts to provide a solution to the identified problems \cite{Hanner:1999fk} have been mostly successful, however, since their production the NTFS has evolved to allow for the specification of fine- and -coarse grained file system permissions \cite{SPE:SPE513}. This brings additional complexity as not only can the standard six permission levels be granted, there is the possibility to create `special permissions' which are constructed from any combination of the possible fourteen permission attributes. 

Microsoft provide a variety of command line utilities \cite{xcalc, accesschk, accessenum} and third-party solutions are also available \cite{scriptlogic} to examine permission allocation. However, the shortcomings of these utilities make none of them serve as a single solution. These shortcomings can be summarised as the inabilities to:
\begin{enumerate}
	\item Show both fine- and coarse-grained permissions.
	\item Examine permissions on multiple folders at once.
	\item Evaluate permissions per user rather than per object.
\end{enumerate}

There is insufficient literature available to suggest that freely available tools have been developed to significantly aid with the administration and reporting of NTFS permissions \cite{russel2003, rusinovich2005, admincomp}, as well as providing detailed information regarding the low-level implementation NTFS access control \cite{carrier2005}. There are few research papers aimed at understanding NTFS access control \cite{wang2006, yue2003} and how it can be improved through better administration \cite{SPE:SPE513}. One author has provided a formal model of NTFS access control, describing fundamentals of rigours implementation \cite{crampton2001}, but there is no indication of the production of any tools that make this available for system administrators.

One paper provides the results for an alternative management interface for NTFS permissions \cite{Maxion2005}. Through careful consideration to human and computer interaction, an application was designed where they could performed administration tasks significantly faster, whilst reducing potential errors. However, the work is restricted to only viewing file system permissions for a single directory at any one time. Since the work was been published, there is no evidence that the tool has been made available in the public domain. Other work includes using novel ways to represent security policies \cite{Reeder2008}. This work is also concerned with temporal aspects of managing file system permissions, whereas the work in this paper is also concerned with providing useful features to aid the quality of the analysis and help to reduce misconfiguration. 

This paper starts by giving a detailed description of how NTFS implements file system permissions, highlighting complexities that result in misconfiguration. A design is then provided, detailing how a software tool can be used to help overcome the complexities, reducing misconfiguration. The next section discusses the functionality of the produced piece of software. This section describes how the functionality can be used to overcome the highlighted complexities by using real-world examples where possible. Finally, we conclude by discussing the beneficial impacts that the solution can bring, and suggest future developments.

\section{NTFS Access Control}
\label{innerworkings}
In this section we describe the inner-workings of the NTFS as regards to permission management. It is necessary to investigate the following aspects to motivate the designed solution.
\subsection{Access control structure}
The NTFS follows in the footsteps of Microsoft's object-oriented approach to implementation. This means that the file system is made up of multiple file and folder objects, and any subject within the operating system (user or process) can request operations on the objects.

\begin{figure}
\centering
\includegraphics[width=0.4\textwidth]{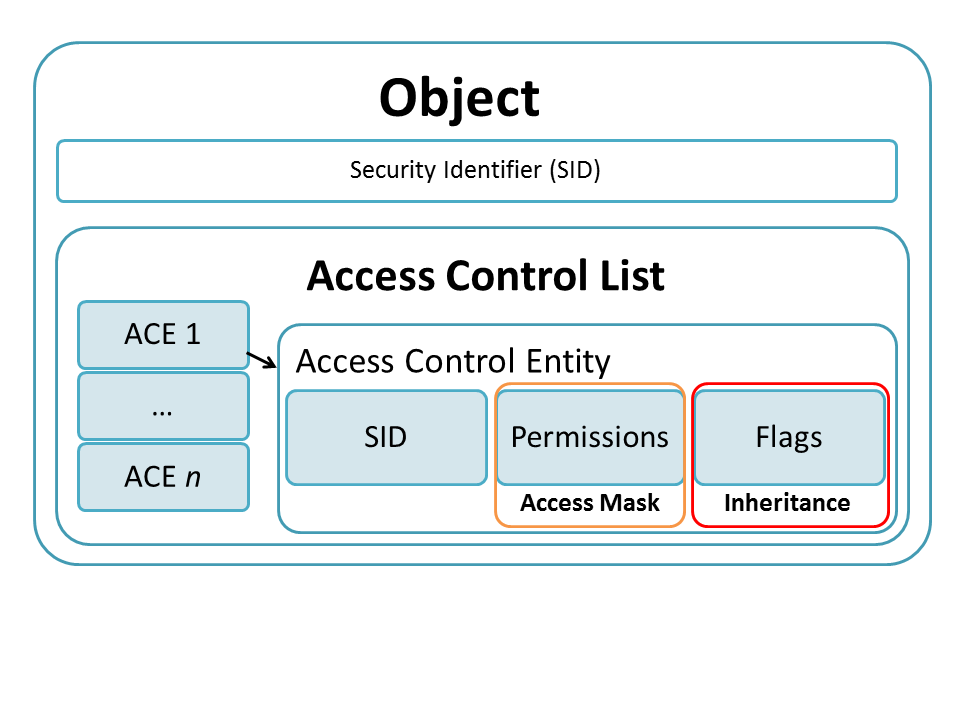}
\caption{Access Control List illustration}
 \label{fig:acl}
\end{figure}

To control access to file system objects, the NTFS implements Access Control Lists (ACLs) by applying an ACL to each object within the file system. Each ACL will contain a Security Identifier (SID) which is a unique key that identifies the owner of the object and the primary associated group. The structure of the ACL is a sequential storage mechanism which contains access control entries (ACEs). An ACE is an element within an ACL which dictates the level of access given to the interacting subject. The ACE contains a SID that identifies the particular subject, an access mask which contains information regarding the level of permissions and the inheritance flags. Figure~\ref{fig:acl} illustrates the logical structure of an ACL and associated ACEs.

\subsection{Access Mask}
An ACE within the NTFS is made up of a combination of fourteen individual permission attributes. The NTFS provides six levels of standard coarse-grained permission that consist of a combination of predefined attributes. It is also the case that NTFS allows for the creation of special coarse-grained permissions which consist of any combination of the fourteen individual attributes\cite{admincomp}.

\begin{table}
\renewcommand{\arraystretch}{1.3}
\caption{Bit mask}
\label{tbl:bittbl}
\centering
\begin{tabular}{lll}
\hline
  Bit / Bit range & Description & Example \\ \\
\hline
    0-15  & Object specific access rights & Read Data, Execute, Append Data \\
   16-22 & Standard security access rights & Delete ACE, Write ACL, Write owner  \\
    23 & Access to ACL & Access System Security  \\
    24-27 & Reserved & n/a  \\ 
    28 & Generic all & $29\cup30\cup21$  \\ 
    29 & Generic Execute & All needed to execute  \\ 
    30 & Generic Write & All needed to write to a file  \\ 
    31 & Generic Read & All needed to read a file \\ 
\hline
\end{tabular}
\end{table}

The access mask is represented by a thirty-two-bit vector. Table \ref{tbl:bittbl} identifies the use of each bit within the vector. It is evident from the table that the standard coarse-grained permissions are represented as follows;

\begin{table}
\renewcommand{\arraystretch}{1.3}
\caption{Standard coarse grained permission bits}
 \label{stsndardattr}
\centering
\begin{tabular}{ll}
\hline
  Coarse-grained level & Set bit(s)  \\ \\
\hline
    Read & bit31\\ 
    Write & bit30\\ 
    List folder contents & bit31 \begin{math} \cup \end{math} bit29\\ 
    Read and execute & bit31 \begin{math} \cup \end{math} bit29\\ 
    Modify & bit31 \begin{math} \cup \end{math} bit29 \begin{math} \cup \end{math} bit30\\ 
    Full control & bit28\\ 
   \hline
   \end{tabular}  
\end{table}

Fine-grained special permissions are represented by using the bits within the range of zero to fifteen. Creating a special permission for most is a very useful feature; however, it can often be a source of confusion as it requires the complete understanding of the authority that each attribute holds \cite{winitpro}.

A good example of having to use special permissions is when you wish to assign a group of users the standard privilege elevation of modify for all the contents of a shared folder. However, creating an ACE with the modify permission on the folder explicitly will result in the user being able to delete the folder itself rather than the child objects (Table~\ref{tbl:bittbl}). To get around this problem we would simply assign the group or user the default permission level of Modify, and then go and modify the permissions' attributes turning it into a special permission so that only sub\-folders and files can be deleted.

\subsection{Propagation and Inheritance}
It is necessary to discuss the different mechanisms behind the way that NTFS permissions can propagate throughout the directory structure. Within the ACL there are two types of ACE; (1) Explicit and (2) Inherited. Explicit entries are those that are applied directly to the objects' ACL, whereas inherited are those that are propagated from their parent object. The type of ACE allows to determine whether the permission was assigned directly to the directory in question (explicit) or if it was inherited from the directory that it resides within (inherited).

This mechanism is controlled by the bit-flag within each ACE as seen in Figure~\ref{fig:acl}. Table~\ref{inheritance} shows the standard three coarse-grained levels of propagation and explains their use.

\begin{table}
\renewcommand{\arraystretch}{1.3}
\caption{Propagation and inheritance}
 \label{inheritance}
\centering
\begin{tabular}{ll p{5cm}}
\hline
    Bit & Name & Use\\\\
\hline
    1 & container inherit ace & Applies the ACE to all the children objects\\ 
    2 & no propagate inherit ace & Propagates the ACE to the child object without bit 1 being set, therefore, stopping propagation at the first level. \\ 
    3 & inherit only ace & The ACE only applies to children objects. (i.e. does not apply to container) \\ 
   \hline
   \end{tabular}  
\end{table}

Furthermore, the creation of fine-grained special file system permissions also allows for the creation of custom fine-grained inheritance rules. Special inherited permissions can be different depending on whether the ACE has the container inherit ace bit flag set which controls whether the ACE is applied to all the children objects or not. The creation of fine- grained propagation rules can easily be overlooked and can ultimately result in the unintended propagation of access.

One of the main difficulties with access propagation with the NTFS is correctly evaluating the effective propagation rules. For a user to view the propagation rules the same situation as viewing the effective permission applies, where the user is required to traverse through the several Windows interface to retrieve the required information as seen in Figure~\ref{fig:explorer}.

\subsection{Accumulation}
\label{accumulation}
Accumulation is the possibility for the subject to receive the effective permission of multiple different policies. This feature is prominent within the NTFS resulting in the possibility for a subject to receive permissions from multiple different ACEs within the same ACL. Furthermore, any subject that interacts with the NTFS can be assigned to any number of groups, which can be entered into the ACE. This means that the user does not have to be directly entered into the ACE, they could simply be a member of the group that is entered.

The policy combination is handled within the operating system by the Local Security Authority Subsystem Service (LSASS). This service combines the permissions together to effectively create the union of all the policies. There are few complexities within permission accumulation due to the structured way in which ACEs are processed. These are:
\begin{enumerate}
	\item Explicit permissions take precedence over inherited permissions.
	\item Explicit deny permissions always take precedence over apply permissions.
	\item Permissions inherited from closer relatives take precedence over relatives.
further away.
\end{enumerate}

\begin{figure}
\centering
\includegraphics[width=0.5\textwidth]{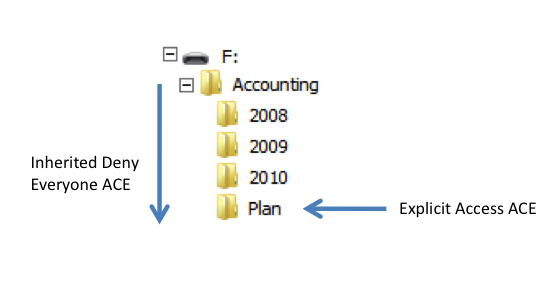}
\caption{Explicit beford inherited demonstration}
 \label{eb4i}
\end{figure}

It might expect that deny permissions always take precedence over apply permissions to ensure that during the policy combination stage the user always operates as the least possible privilege elevation. However, the first point regarding explicit permissions taking precedence over inherited permissions can result in a situation where an inherited deny permission is never reached. Considering the folder structure in Figure \ref{eb4i}, where the folder Accounting has an explicit deny permission for the Everyone group, which is set to propagate to all its children. This means that all the sub\-folders to the Accounting folder will receive an inherited deny Everyone ACE. If the case was to arise, like in this example, where a single user now requires access to the Plan folder, an explicit ACE to allow access could be entered. Now when the user visits the Plan folder, the LSASS would process the explicit allow permission first and allow for it to take precedence over any other permission. This goes against a fundamental aspect of policy combination to ensure that a deny permission is never ignored. If the case where a user is able to ignore a deny permission to receive access was to either intentionally or unintentionally arise, the system administrator needs to be made aware of this situation.

To summarise, the precedence hierarchy for policy accumulation is as follows:

\begin{enumerate}
	\item Explicit deny.
	\item Explicit allow.
	\item Inherited deny.
	\item Inherited allow.
\end{enumerate}

In addition to the explicit permissions taking precedence over inherited permissions, inherited permissions that of closer distance to the invoked object will take precedence over more distant relatives. For example, a folder's inherited permissions will take precedence over those from their grandparent.

Accounting for permission accumulation has currently been made possible by using the standard Windows Explorer feature of displaying the effective permission. This feature allows for the user to enter a specified user or group and the effective permission that they hold on that specific directory will be displayed. Unfortunately, performing this evaluation on several folders soon becomes infeasible.

\subsection{Group Membership}

A fundamental aspect of access control within the NTFS is that of group membership. A subject (group, user or process) that interacts with the file system can be a member of any group. This means that permissions can be inherited from any of the associated groups if they are entered within any ACL. Subjects, in this case users, will often be grouped together by (separation of duty) to make management easier, and as Hanner, 1999 \cite{Hanner:1999fk} identifies, understanding effective file permissions can become significantly more complex by group association. To correctly evaluate a user's effective permissions you would have to know which groups they are a member of. We should note that this is not directly related to the mechanism of how NTFS implements access control, it is an unavoidable component of how Microsoft allows for users, groups and processes to be managed by group association.

\section{Novel Solution}

This section describes the design of a solution based on the NTFS's inner-workings which can cause the identified administrative complexities as seen in Section \ref{innerworkings}.

\subsection{Coarse- and Fine-Grained Permissions}
\label{compression}
As previously described, the NTFS allows for the standard set of coarse permissions, but also allows for the creation of special fine-grained permissions. 

An alternative method of display, special permissions could be displayed by a character-to-attribute representation. This way a string can be constructed to display the full granularity of the permission by only using little space.  For example, if a special permission was constructed to have the attributes enabled: 
\begin{enumerate}
	\item Read (R).
	\item Write (W).
	\item Delete sub\-folders and files (Dc).
	\item Read permissions (Rp).
	\item Change permissions (Cp).
\end{enumerate}

Using the character-to-attribute would results in the production of the string `R-W-Dc-Rp-Cp'. After some time the user would become accustom to this relationship and the key would no longer be required.

\subsection{Multiple Folders}
\begin{algorithm}
 \SetAlgoLined\DontPrintSemicolon
\KwIn{Initial directory  $d$}%
\KwIn{Set of ACEs to be filtered out $F=(f_1, f_2, f_3, \ldots, f_n)$}
\KwOut{Set of ordered directories and ACEs $P = (d_1, (p_1, p_2, p_3, \ldots,p_n ))$ where $d_n$ is the directory and $p_n$ are the permission entries for that directory.}\
  \SetKwFunction{algo}{algo}\SetKwFunction{proc}{proc}
  \SetKwProg{myalg}{Algorithm}{}{}
  \myalg{\algo{}}{
  \nl  $P \gets  $ \proc{d}\;
  \nl \KwRet}{}\;
  \setcounter{AlgoLine}{0}
  \SetKwProg{myproc}{Procedure}{}{}
  \myproc{\proc{directory d}}{
 	$pACL \gets d(ACL)$\;	
	\lForEach{subdirectory $c$ of $d$}{\;
		$cACL \gets c(ACL)$\;
		\If{$cACL\; != pACL$}{
			\lForEach{ACE $a$ in $cACL$}{\;
				\If{$a \not\in F$}{
					\uIf{$isSpecial(a)$}{
						$p \gets compress(p)$\;
					}
					\Else{
						$p \gets a$
					}
					$P \gets (c, p)$\;
					\proc{c}
					}
		}	}
}}
\caption{Depth-first recursive directory search, analysing and filtering security permissions. }
\label{algo:search}
\end{algorithm}

It has previously been identified that Windows Explorer allows for the examination of an objects' ACL, however, it is often the case that evaluating multiple ACLs is necessary. A useful way to view multiple ACLs would be to allow the  examination of a whole directory structure simultaneously. This would provide the means to also examine how the propagation and inheritance aspects of the ACLs are interacting. Algorithm~\ref{algo:search} describes the recursive depth-first examination search technique that has been implemented for analysing the permissions of multiple folders. This algorithm traverses the directory structure, analysing each directories permissions. In each analysis, the algorithm evaluates whether:
\begin{enumerate}
	\item It is necessary to display the current ACL to the user based on whether it is different from the parent's ACL.
	\item Each ACE in the ACL contains a special permission.
	\item Report the ACE to the user, displaying the level of permission.
\end{enumerate}

\subsection{Compression} 
As seen on line 9 of Algorithm~\ref{algo:search}, a compress function is called if a special permission is identified. This compress function performs the character-to-attribute mapping as described in Section~\ref{compression}. In this method, an enumerated type is used for changing the permission attributes to the associated character.

\subsection{Filtering} 
Filtering of groups is easily performed as shown on line 7 of Algorithm~\ref{algo:search} where a check is made to ensure that the current ACE $a$ is not present in the set of groups to filer $F$. This provides the facility to filter for multiple user or group objects, therefore removing excess information.

\subsection{Per User View}
When performing a per user search of the file system,  Algorithm~\ref{algo:search} is used, however, line 7 is substituted with a condition to check that the ACE in question is the one that is being searched for ($a \in F$). This means that all groups and user objects are excluded if they are not represent in the filter list. When viewing per user, the filer list contains the user or group that the user wants to analyse. 

\subsection{Accumulation}
Algorithm~\ref{algo:search} identifies provides a search strategy that can report the file system permissions for an entire directory structure, whilst considering compression and filtering. Although the returned permission information is what is visible in the ACE, it might not be the user's effective permission as no consideration to permission accumulation as described in Section~\ref{accumulation} is taken. Algorithm~\ref{algo:accum} provides an alternative method where the search concentrates on calculating the effective permission that the user and or group hold. Algorithm~\ref{algo:accum} shows an algorithm that can be used to store the explicit $ex$ and inherited $in$ permissions based on the inheritance and propagation. This algorithm considers both the inheritance and deny hierarchies. For speed purposes the algorithm can identify deny permissions and stop the algorithm from continuing the examine the ACL. Line 16 shows that once the explicit and inherited permissions have been identified a function is then called to calculate the effective permission. In this algorithm $calculatedEffective(explicit,inherited)$ represents a native Microsoft .NET command that is able to return the effective permission. Using this native method ensures that the correct effective permission is reported.

\begin{algorithm}
 \SetAlgoLined\DontPrintSemicolon
\KwIn{Initial directory  $d$}%
\KwIn{Initial group or user  $u$}
\KwOut{Set of ordered directories and ACEs $P = (d_1, (p_1, p_2, p_3, \ldots,p_n ))$ where $d_n$ is the directory and $p_n$ are the permission entries for that directory.}\
  \SetKwFunction{algo}{algo}\SetKwFunction{proc}{proc}
  \SetKwProg{myalg}{Algorithm}{}{}
  \myalg{\algo{}}{
  \nl  $P \gets  $ \proc{d}\;
  \nl \KwRet}{}\;
  \setcounter{AlgoLine}{0}
  \SetKwProg{myproc}{Procedure}{}{}
  \myproc{\proc{directory d}}{
 	$pACL \gets d(ACL)$\;	
	\lForEach{subdirectory $c$ of $d$}{\;
		$cACL \gets c(ACL)$\;
		\If{$cACL\; != pACL$}{
		$ex = \emptyset$,
		$in = \emptyset$\;
			\lForEach{ACE $a$ in $cACL$}{\;
				\uIf{$isExplicitDeny(a)$}{
				 	$P \gets (c, a)$\;
					break\;
				}
				\Else{
				\uElseIf{$isExplicitAllow(a)$}{
					 $ex \gets a$\;
				}
				\uElseIf{$isInherited(a)$}{
					 $in \gets a$\;
				}
				$P \gets (c, calculatedEffective(ex,in))$\;
				}	
		}	}
}}
\caption{Depth-first recursive directory search, returning the effective permission of a specified user or group.}
\label{algo:accum}
\end{algorithm}

\subsection{Group Membership}
User and group membership is fundamental mechanism that allows users to inherit file system permissions from group objects. A simple recursive method can be used to examine a user or groups membership. There are two possible directions in which the group membership can be analysed. The first is to examine which groups an object is a member of. This is where a search is performed to recursively report which groups a user or group is a member of. The second method is the members of displaying a user or groups members.  This is where a recursive search is performed to reporting on a groups members.

\section{Developed Solution}
The developed software-based tool is programmed in C\# .NET 3.5 with the use of the Microsoft Management Console (MMC) System Development Kit (SDK) to produce a MMC SnapIn application. The motivation behind making the application run in the MMC was to bring consistency with other Microsoft management tool, therefore, making the software self-intuitive for the users. 

The software runs under the credentials of the executing user, therefore, only receiving access to view file system permissions that they have been assigned to. The software runs in real-time, processing the desired ACLs upon request. This means that the software requires only a minimal amount of installation, and does not require an additional database to store permission entries. The overheads caused by the application on both the host machine and any interacting file servers are very small and do not affect normal performance at all.

In this remaining of this section, the provided functionality is discussed, using examples where possible.

\subsection{Application Layout}

\begin{figure}
\centering
\includegraphics[width=3.6in]{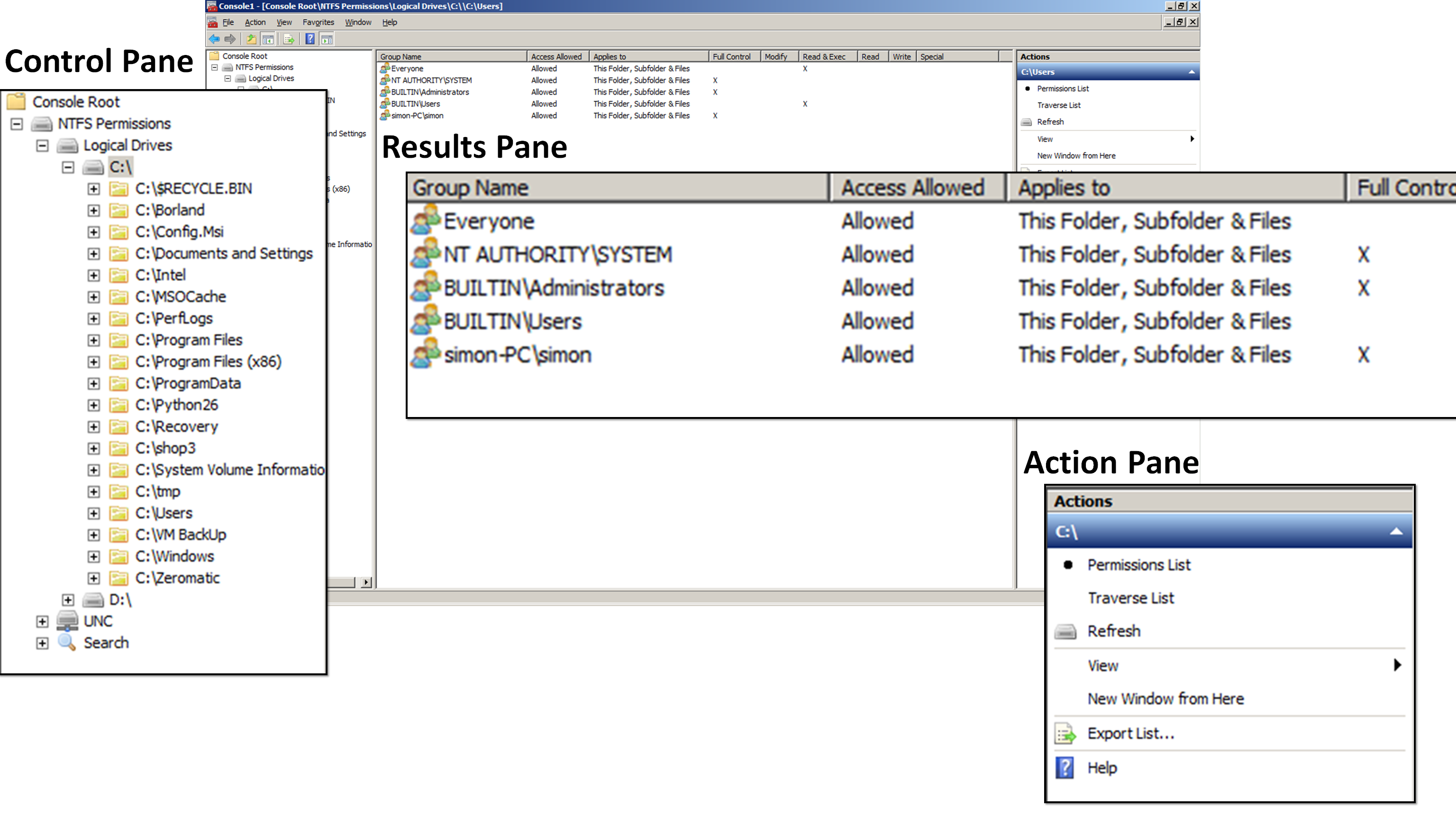}
\caption{Developed MMC Application}
\label{fig:explorer}
\end{figure}

As seen in Figure~\ref{fig:explorer}, the interface has three main sections. Firstly on the left is the control pane. The control pane is where the user can see all the physical and remote mounted NTFS volumes. The user is able to browse the folder structure of all local and remote drives in a Windows standard hierarchical tree view. In addition, any effective permission searches that the user performs will be listed here. The middle pane is where the associated results from the item selected within the control pane are displayed. On the right is the action pane. This pane contains functionality associated with each of the items selected within the control pane that can affect the contents of the results pane. 

The results pane shows the ACL for the specified local or remote drive, providing that the executing user has permission to view the ACL. This pane contains the same ACL information as present in the Windows Explorer interface. The ACEs are classified into the standard NTFS sets although List Folders is not classed as a set because the permission is the same as Read \& Execute, just the propagation is different, which is correctly displayed.

\subsection{Coarse- and Fine-Grained Permissions}
As described in the design, the application does have a different way of representing special permissions. To allow the user to easily and correctly see the fine-grained permissions the special permissions are displayed as a hyphen separated character string, where each character is associated with a different special permission attribute. 

\begin{figure} 
\centering
\includegraphics[width=2in]{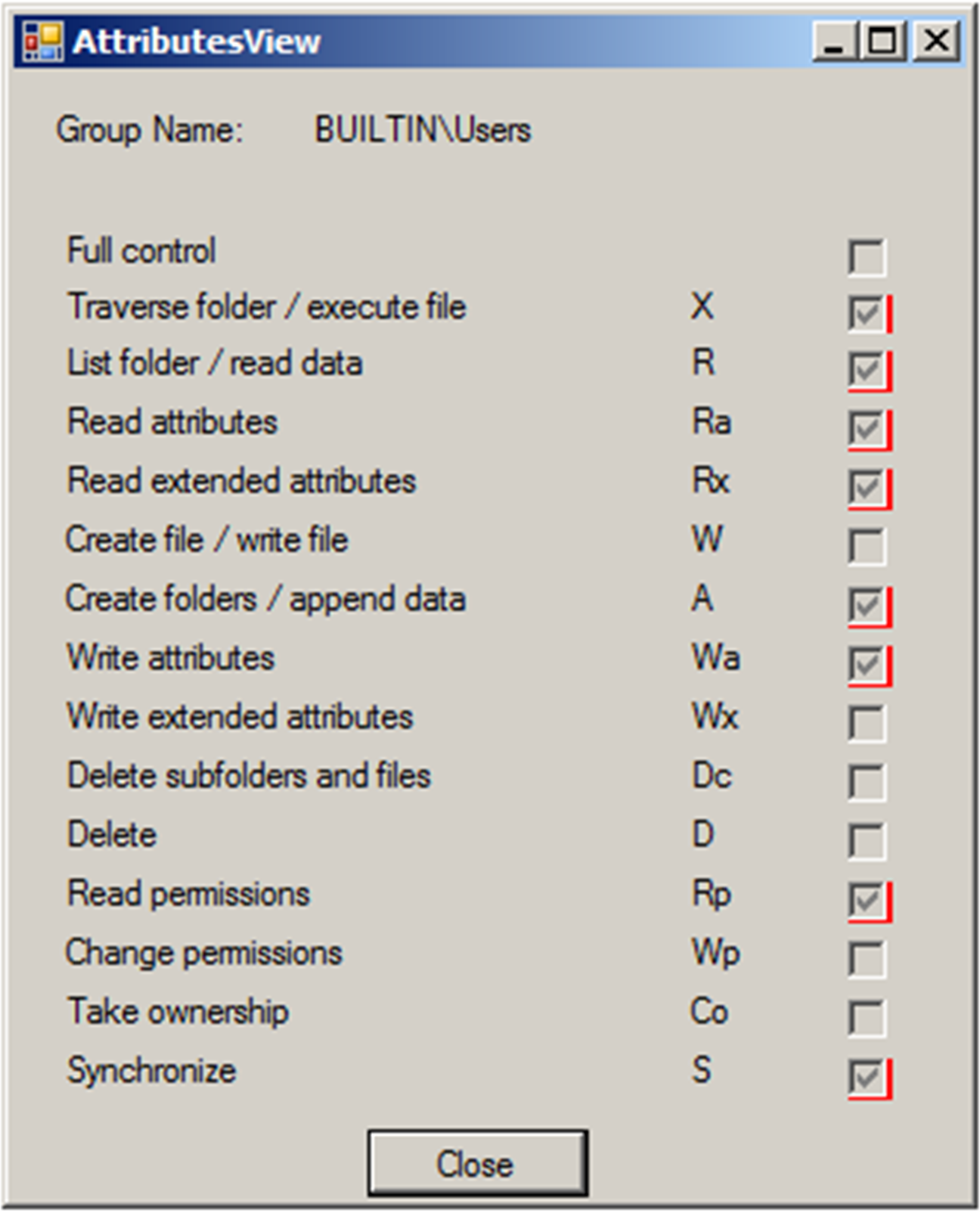}
\caption{Developed MMC Application}
\label{fig:coarse}
\end{figure}

As shown in Figure~\ref{fig:coarse} the group `BUILTIN\textbackslash{}Users'  has a special permission entry that is displayed by the hyphenated character string. On further inspection of this permission it is possible to view the character-to-attribute relationship, which is also displayed in Figure~\ref{fig:coarse}. After using the application we might start to remember the character-to-attribute relationship, meaning that we do not need to inspect the special permission, therefore, further speeding up the process of reporting fine-grained special permissions. The results pane also shows information regarding whether each permission (ACE) is an allow or deny permission, and also the propagation level of each of the ACE entries.

\subsection{Traversal View and Custom Filter}
Another highlighted problem was difficulties within trying to view the ACL for multiple folders at any one time. The developed application avoids this issue by firstly allowing a user to simply traverse the file system in the control pane to view the ACL for a single folder, and secondly, allowing the user to view the ACLs for a whole directory in one traversal view. To reduce the quantity of displayed information and help display what is useful to the user, by default the traversal view will only show the ACL for a folder that is not the same as its parents'. A custom filter has also been implemented so that the user can select groups and users that they do not wish to include in the traversal view. 

\begin{figure} 
\centering
\includegraphics[width=3.5in]{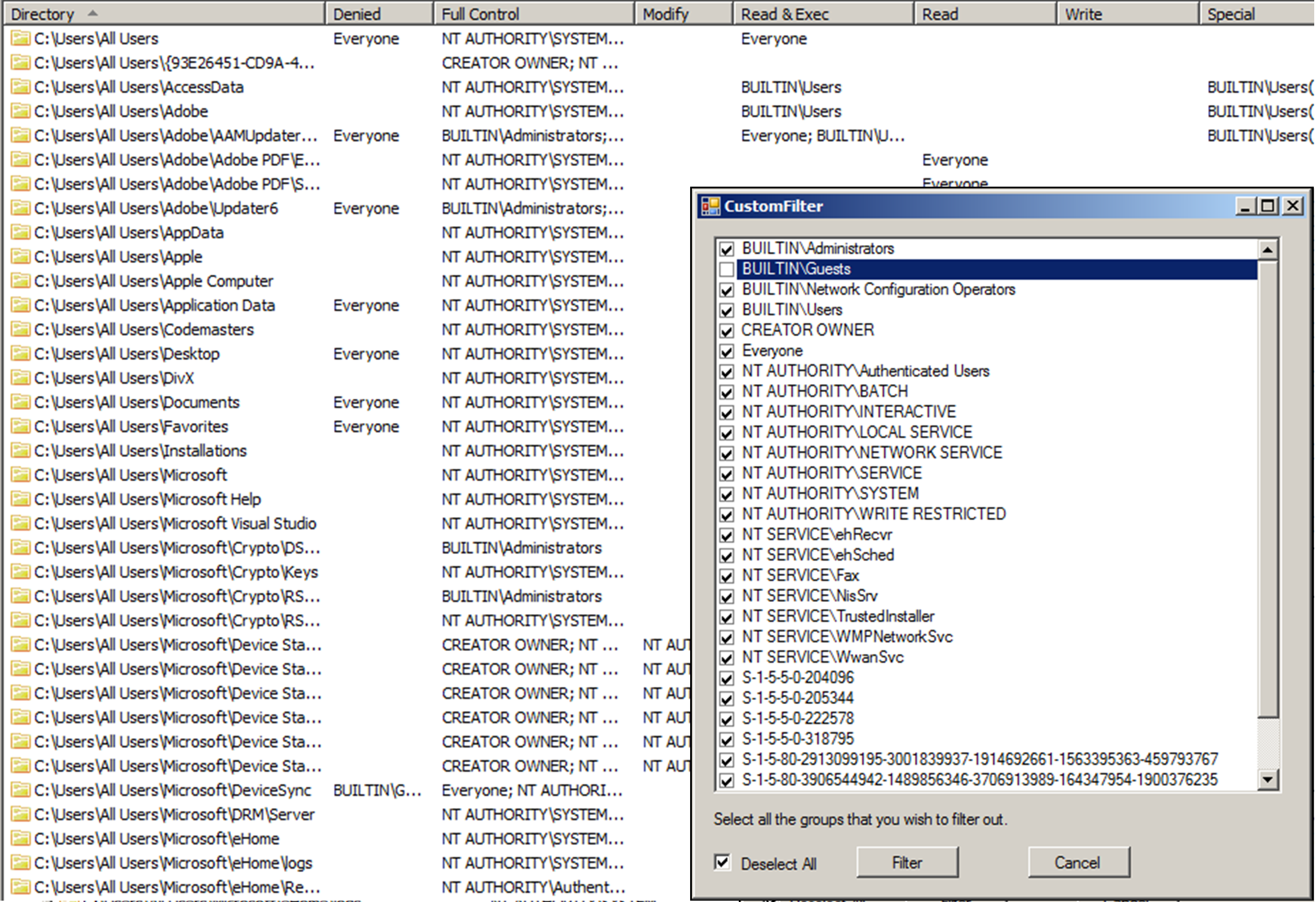}
\caption{Traversal view with custom filter}
\label{fig:traversal}
\end{figure}

Figure~\ref{fig:traversal} shows the results pane when the traversal function is applied to the local folder C:\textbackslash{}Users. The illustration also shows the filter interface where the user can select groups that they wish to remove from view. The traversal view also displays both fine- and coarse-grained permissions in the same way as the individual view where the permissions are classified as the standard or special sets.

\subsection{Permission Accumulation}
\begin{figure} 
\centering
\includegraphics[width=3.5in]{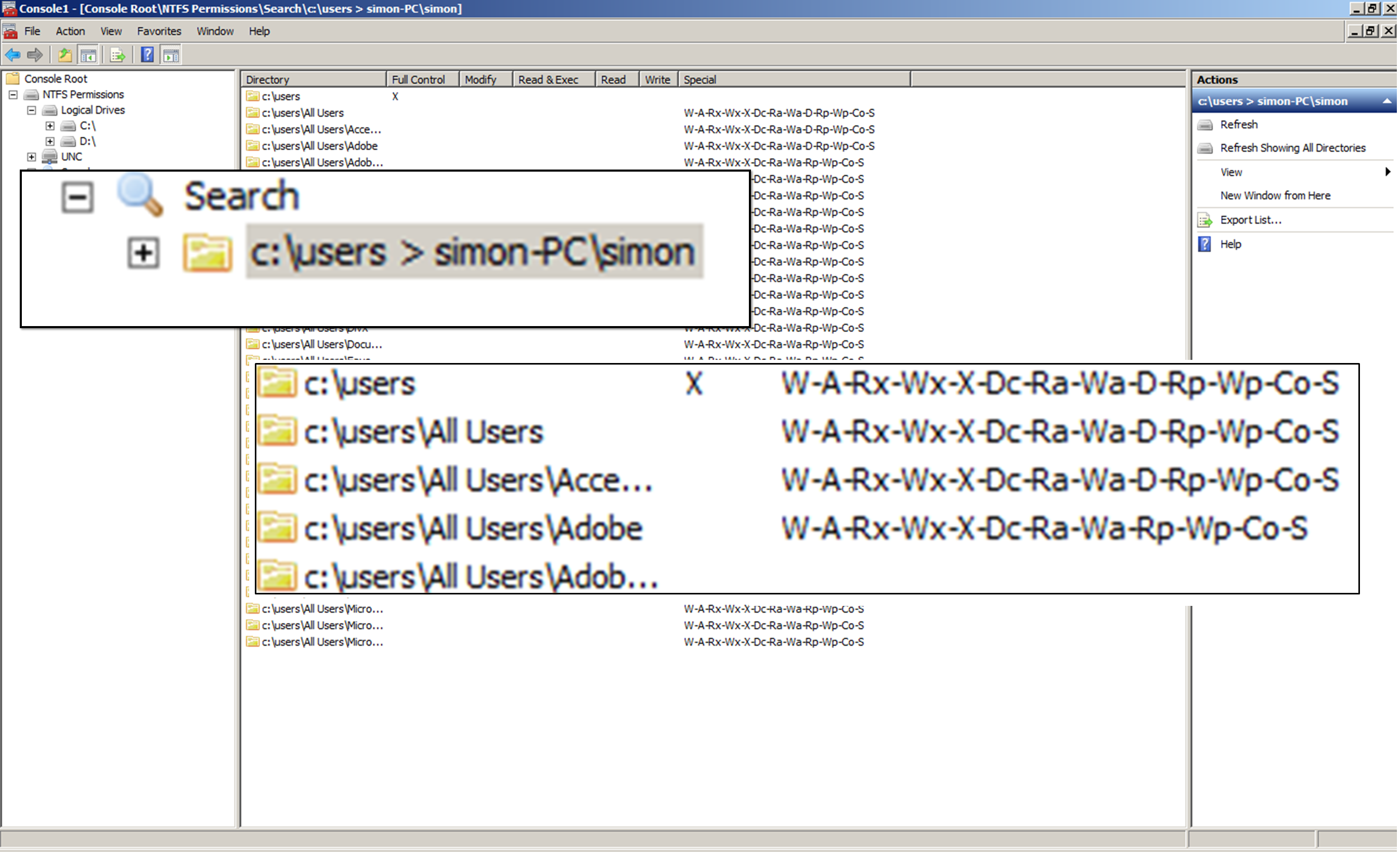}
\caption{Permissions accumulation search results}
\label{fig:search}
\end{figure}

Policy combination can be one of the most time consuming aspects of the NTFS when trying to evaluate the permission that a subject holds on any given location. As described earlier, accumulation of deny and access permissions, group membership as well as consideration to the ACE processing hierarchy results in several complication factors to the evaluation. The developed application has a built-in search feature to show the exact effective permissions for a given subject on the selected location. Figure~\ref{fig:search} shows the interface after performing a custom search for the user `simon-PC\textbackslash{}simon' on the directory `C:\textbackslash{}User'. The same logic applies when performing a search where only permissions that differ from their parent object are displayed by default, and special permissions are displayed using the hyphenated character representation.

\section{Conclusions}
We began by examining in detail the workings of access control within the NTFS to highlight the potential causes of complexity, which could ultimately lead to unintended access. Next, we discussed the common usability problems that can be experienced when examining NTFS permissions. Following this, we developed a Microsoft Management Console SnapIn application to provide a new way of examining NTFS permissions that can help overcome the identified complexities. We believe that our study and software solution helps to improve file system security by providing an intuitive, efficient and thorough method for permission examination. 

This paper provides a contribution to system administrators by aiding them with permission examination and allocation. The requirement to provide a software-based tool to overcome the identified complexities can be established from the in excess of four thousand downloads the tool has received since production. This shows that NTFS administrators are actively seeking support for their duties. In addition to the number of downloads, the tool has also received promotion through a rated software site\cite{softsea} and a useful list of system administration tools\cite{101}. This emphasises how requirement for such tool.

\section{Future Scope}

Future work involves allowing for the user to modify file system permissions once a problem has been identified. Another possibility is a software tool that can automatically identify configuration problems and suggest intelligent solutions.

\section{Acknowledgement}
The authors would like to express great thanks to Michele Puri of the European University Institution for passing on vast amounts of knowledge regarding the implementation and administration of NTFS permissions within a large organisation. Thanks should also be expressed to Alan Radley and Malcolm Merrington of the University of Huddersfield for providing additional insight to the problems and for testing the developed software. 



\bibliographystyle{IEEEtran}
%


\end{document}